\documentstyle[preprint,aps]{revtex}
\tightenlines
\begin{document}
\draft
\preprint{\vbox{To be submitted to Physical Review {\bf C}
\hfill  \hfill USC(NT)-97-05}}
\tolerance = 10000
\hfuzz=5pt
\title{PCAC constraints on pion production/absorption within \\ 
nonrelativistic nuclear dynamics}
\author{V. Dmitra\v sinovi\' c$^{1}$, and T. Sato$^{2}$}
\address{{\rm 1.} Department of Physics  and Astronomy,\\
University of South Carolina, Columbia, SC 29208 \\ 
{\rm 2.} Department of Physics, Osaka University,\\
{Toyonaka, Osaka 560-0043, Japan}} 
\date{\today}
\maketitle
\begin{abstract}
We show the necessity of two-nucleon axial currents and 
associated pion emission/absorption operators 
for the partial conservation of the axial current (PCAC)
nuclear matrix elements with arbitrary nuclear dynamics 
described by a nonrelativistic Schr\" odinger equation.
As examples we construct such nonrelativistic axial two-body
currents 
in the linear- and the heterotic ($g_A = 1.26$) sigma models, 
with an optional isoscalar vector $(\omega)$ meson exchange. 
The nuclear axial current matrix elements obey PCAC
only if the nuclear wave functions used in the calculation are 
solutions to the Schr\" odinger equation with the static
one-meson-exchange potential constructed in the respective 
(sigma) model. The same holds true for the nuclear pion 
production amplitude, since it is proportional to the divergence 
of the axial current matrix element, by virtue of PCAC.
Thus we found a new consistency condition between the 
pion creation/absorption operator and the nuclear Hamiltonian. 
We present examples drawn from our models and discuss the 
implication of our results for one-pion-two-nucleon processes.

\end{abstract}
\pacs{PACS numbers: 25.30.-c, 25.80.Hp, 25.10.+s}
\widetext

\section{Introduction}

In an earlier publication \cite{ax96} a systematic study of 
axial current (partial) conservation was begun in the 
Bethe-Salpeter (BS) approach to nuclear bound states. 
There it was found that partial conservation of axial current 
(PCAC) puts constraints not only 
on the form of the axial current operator, but also on the
nuclear wave function, by way of fixing the 
``potential" entering the nuclear BS equation. Since the pion
production/absorption amplitude is an integral part 
of the conserved axial current nuclear matrix element,
the same constraints are imposed on it as well.
It stands to reason that the same kind of constraint will
carry over into the nonrelativistic (NR) formalism.
\footnote{The idea that PCAC determines the pion-nuclear
production operator has been around at least since the work 
of Blin-Stoyle and Tint \cite{bst67}. But the notion that the
nuclear wave functions entering the same pion production
amplitude are also constrained by the PCAC appears to be new.} 
In this paper we extend the work in Ref. \cite{ax96} to the 
nonrelativistic description of the nucleus. We do not make 
a direct nonrelativistic reduction of the BS equations and 
amplitudes from Ref. \cite{ax96}, 
for they are well known not to have a good NR limit, but rather
use this reference 
as a guide to developing the corresponding 
result within the Schr\" odinger equation approach. We shall
show that even at this nonrelativistic level there are
differences between the various versions of the $\sigma$ model
used in Ref. \cite{ax96}. 

The aforementioned PCAC consistency condition between the 
(elementary) pion production mechanism
and the NN potential determining the nuclear wave equation 
is a novel feature promising to introduce 
a higher level of logical coherence into present day 
calculations of pion production on nuclear targets. As an
example one may consider the $pp \rightarrow \pi^{0}pp$ and
$pp \rightarrow \pi^{+}d$ reactions. 
There, in particular, the $\pi$-production mechanism (operator) 
has often been considered without any reference to the nuclear 
wave functions \cite{bst67,kr66}. 
Several models involving scalar $\sigma$-meson-exchange 
pion production ``current", besides the $\pi$-exchange one, 
have been proposed \cite{hor93,riska93}. 
We shall show here exactly which $\sigma$-meson-exchange 
$\pi$-production operator must be included and how when the
two-nucleon potential contains a one-sigma-exchange term, and 
{\it vice versa}. It turns out that this $\pi$-production 
operator is rather sensitive to the details of the sigma model 
used.
We also analyze the $\omega$-meson-exchange $\pi$-production 
operator required by the presence of the $\omega$-meson-exchange 
potential.

This paper falls into five sections. After the Introduction,
in Section~II, 
we define the context of our analysis and prove our 
most general result: the necessity of two-body axial currents
for interacting two- (or more) particle systems obeying 
nonrelativistic quantum mechanics.
In section~III we construct two-nucleon axial currents that
respect PCAC at the level of {\it nuclear} matrix elements, 
starting from three underlying relativistic chirally symmetric 
meson-nucleon 
model Lagrangians.
In Section~IV we present the three corresponding sets of 
two-nucleon $\pi$-production operators as an example of the 
main proposition and we discuss the results. 
In Section~V we summarize and draw the conclusions.

\section{Partial conservation of nuclear axial current}

The notion of PCAC is both historically and conceptually 
the foundation of chiral symmetry in hadronic interactions.
Modern implementations of this symmetry, such as the nuclear
chiral perturbation theory ($N\chi PT$), however, do not 
emphasize that point. 
In the following we shall try and present another viewpoint
of PCAC that will bear significance to nuclear physics in
general and pion-nuclear processes in particular.
In short, PCAC states that the hadronic axial 
current $J_{\mu 5}^{a}$ must satisfy the following 
continuity equation
\begin{eqnarray}
\partial^{\mu} J_{\mu 5}^{a}
&=&  
- f_{\pi} m_{\pi}^2 \Pi^{a} ~,\
\label{e:lsdiv}
\end{eqnarray}
or equivalently
\begin{eqnarray}
\bbox{\nabla} \cdot {\bf J}^{a}_{5}({\bf R}) + 
{\partial \rho^{a}_{5}({\bf R}) \over \partial t} = - 
f_{\pi} m_{\pi}^2 \Pi^{a}({\bf R}) , 
\label{e:ntcont}
\end{eqnarray}
where $\Pi^{a}$ is the (canonical) pion field operator. 
In the quantum mechanical framework this can be written as an
equation relating the divergence of the three-current 
and the commutator of the Hamiltonian and the axial charge 
density:
\begin{eqnarray}
\bbox{\nabla} \cdot {\bf J}^{a}_{5}({\bf R}) 
+ i \left[~H,~\rho^{a}_{5}({\bf R})
\right] = - f_{\pi} m_{\pi}^2 \Pi^{a}({\bf R})~. 
\label{e:qmcont}
\end{eqnarray}
This equation is a consequence of the (exact) Heisenberg equations
of motion. Now we specialize to nonrelativistic nuclear physics 
by limiting 
ourselves to that subspace of the complete Hilbert space that 
contains at most two (real) nucleons interacting by exchanging 
one (virtual) meson at a time.
There the total Hamiltonian of the nucleus $H$ is the
sum of the kinetic and potential energies $H = T + V$ of the
nucleons, and
the total axial current ${\bf J}^{a}_{5}({\bf R})$ consists of 
one- and two-nucleon parts. We assume that the axial charge 
density $\rho^{a}_{5}$
\footnote{
In the following we shall drop the ``index variable" ${\bf R}$ 
in the current and charge operators, except when necessary to 
avoid confusion.} is well approximated by its one-nucleon part
\footnote{
The subleading (relativistic) 
correction to the axial charge density operator contains 
two-body terms that modify the following analysis somewhat, 
but cannot change its main conclusion, due to an intrinsically 
different tensor structure of the one- and two-body operators.} 
$\rho_{5, {\rm 1-b}}^{a}$.
This assumption agrees with - indeed it follows from - our
fundamental assumption of nonrelativistic nuclear dynamics.
This means that all (relativistic) operators are expanded in 
powers of $1/M$; this provides a convenient book-keeping device
in what follows.
As shall be shown below, the (partial) continuity equation
(\ref{e:qmcont}) 
strongly constrains (the longitudinal part of) the nuclear axial 
current, and in particular its two-nucleon, or meson-exchange 
part. 

We may break up the axial current conservation equation 
into one- and two-body parts without loss of generality. 
The divergence of the complete
one-body current equals $-i$ times the commutator of the kinetic 
energy $T$ and the one-body axial charge density
\begin{eqnarray}
\bbox{\nabla} \cdot {\bf J}^{a}_{5} ({\rm 1-body}) = 
- i \left[~T,~\rho_{5} ({\rm 1-body})\right]
- f_{\pi} m_{\pi}^2 \Pi^{a}({\rm 1-body}) ~,
\label{e:1bdiv}
\end{eqnarray}
is of ${\cal O}(M^{-2})$, i.e., zero to leading order in $1/M$, 
due to similar momentum dependencies of the kinetic energy $T$ 
and the axial charge density $\rho^{a}_{5} ({\rm 1-body})$
operators, as well as to the absence of non-diagonal isospin 
operators from $T$.
Therefore, the test of conservation of the complete nuclear 
axial current is whether or not the potential $V$ commutes 
with the one-body axial charge density. It turns out that, 
due to the momentum operator inside of 
$\rho_{5}^{a}({\rm 1-body})$, only 
a completely trivial, viz. a spatially everywhere constant 
potential commutes with the axial charge.
In nuclear physics, therefore, one {\it always} needs a two-body axial current 
axial current ${\bf J}^{a}_{5}({\rm 2-body}) = 
\sum_{j < k}^{A} {\bf J}_{5, (jk)}({\rm 2-body})$ to compensate 
for the temporal change of the axial charge density. 

To show this formally we note that 
in general there are two possible sources of the 
non-commutativity of meson exchange potential $V$ and the 
one-body axial charge density $\rho^{a}_{5}$:
(i) non-commuting isospin factors; and (ii)
non-commuting spin-spatial factors, as can be seen from the
identity
\begin{eqnarray} 
\bbox{\nabla} \cdot {\bf J}^{a}_{5}({\rm 2-body})
&=& - i \left[V,~\rho^{a}_{5}\right]
- f_{\pi} m_{\pi}^2 \Pi^{a}({\rm 2-body}) 
\nonumber  \\
&=& 
{i \over 2} \sum^{A}_{i=1} \sum_{j<k}^{A}  
\left[{\tau^{a}_{(i)} \over {2 M}} , I_{(jk)} \right]
\left\{ {\cal V}_{(jk)}, 
\left\{\bbox{\sigma}_{(i)} \cdot \bbox{\nabla}_{(i)}, 
\delta({\bf R} - {\bf r}_{(i)}) \right\}
\right\}
\nonumber \\
&-& {i \over 2} 
\sum^{A}_{i=1} \sum_{j<k}^{A}  
\left\{{\tau^{a}_{(i)} \over {2 M}} , I_{(jk)} \right\}
\left[ {\cal V}_{(jk)}, 
\left\{\bbox{\sigma}_{(i)} \cdot \bbox{\nabla}_{(i)}, 
\delta({\bf R} - {\bf r}_{(i))} \right\}\right]
\nonumber \\
&-& 
f_{\pi} m_{\pi}^2
\Pi_{\pi}^{a}({\rm 2-body})  ~, \
\label{e:2bdiv}
\end{eqnarray}
where the two-body potential 
$V_{(jk)} = I_{(jk)}{\cal V}_{(jk)}$ is a product of its isospin
$I_{(jk)}$ and spin-spatial ${\cal V}_{(jk)}$ parts.
Since 
\begin{eqnarray}
\left[{\cal V}_{(jk)},\left\{\bbox{\nabla}_{(i)}, 
\delta({\bf R} - {\bf r}_{(i)})\right\}\right] \propto
\delta({\bf R} - {\bf r}_{(i)}) \delta_{ij}
\left[\bbox{\nabla}_{(j)}, {\cal V}_{(jk)}\right]
~ \neq 0~,
\label{e:neq0}
\end{eqnarray} 
and the isospin anticommutator 
$\left\{\tau^{a}_{(i)} , I_{(jk)} \right\} \neq 0$
does not vanish for at least one value of $a=1,2,3$, we see 
that the next-to-the-last line in Eq. (\ref{e:2bdiv}) does not 
vanish, and thence 
{\it the commutator of the potential and the axial 
charge density does not vanish for any potential} 
${\cal V}_{jk}$, except the trivial one, i.e., a 
constant: ${\cal V}_{jk} = {\rm const.}$
\footnote{
This argument does not furnish a formal proof so long 
as all possible spin and isospin operators have 
not been examined for accidental vanishing of their
anticommutators with $\bbox{\sigma}_{(i)}$ and $\tau^{a}_{(i)}$, 
respectively. Our proposition has been confirmed in all 
cases that we have studied so far, which cases constitute some
of the most important parts of the nuclear NN potential.}.
In other words, {\it axial meson exchange currents (MEC) 
${\bf J}^{a}_{5}({\rm 2-body})$
are always necessary, so long as two nucleons interact and 
their dynamics can be described by Quantum Mechanics}
\footnote{
This does not mean, however, that we know how to construct axial 
MECs that obey PCAC for arbitrary NN potentials. Presently we
know how to calculate only those axial MECs that are related to
potentials that are based on chiral Lagrangian models. Methods
ordinarily used for electromagnetic (EM) MECs do not seem to
work here, cf. Ref. \protect{\cite{ax96}}.}.  
The same conclusions, of course, hold in nonrelativistic (NR) 
quark models, i.e., axial two-quark currents are always 
necessary in NR quark models, as well.
It is the compelling nature of this argument that makes it
distinct from previous arguments along similar lines
\cite{rho79,ffo76}. 

This is perhaps a somewhat surprising result in view of the
fact that the EM current conservation does {\it not}
require MECs for many parts of the nuclear potential, e.g. for 
the contributions from the exchange of neutral mesons. Hence it 
will be our task to construct axial MECs associated with the 
exchange of the most important, i.e., the lightest, mesons. The
$\pi$-exchange axial current has been known for some time
\cite{oht79,sauer92}, so we shall not repeat its derivation
here. In this paper we shall concentrate on the isoscalar scalar
$(\sigma)$ and vector $(\omega)$ mesons. The isovector vector
$(\rho)$ meson cannot be introduced in a model-independent way,
so we leave it for another occasion. 

To reveal the necessity of consistency between the nuclear wave 
functions and the nuclear pion production/absorption amplitude we 
must remember that the operator equation (\ref{e:qmcont}) 
describing PCAC is just a shorthand for the same statement 
about {\it all} axial current nuclear matrix elements 
$\langle J_{\mu 5}^{a}\rangle_{fi}$. In momentum space the 
Heisenberg equations of motion lead to
\begin{eqnarray}
q^{\mu} \langle J_{\mu 5}^{a} \rangle_{fi} 
&=& 
{\bf q} \cdot \langle {\bf J}_{5}^{a} \rangle_{fi} -
(E_{f} - E_{i}) \langle \rho^{a}_{5} \rangle_{fi}  
\nonumber \\
&=& 
{\bf q} \cdot \langle {\bf J}_{5}^{a} \rangle_{fi} -
\langle \left[H,~\rho^{a}_{5}\right] \rangle_{fi}  
\nonumber \\
&=& 
i \left({f_{\pi} m_{\pi}^2 \over{q^{2} - m_{\pi}^2}} \right) 
\langle \Gamma_{\pi}^{a} \rangle_{fi}  
\nonumber \\
&\simeq& 
- i
\left({f_{\pi} m_{\pi}^2 \over{{\bf q}^{2} + m_{\pi}^2}} \right) 
\langle \Gamma_{\pi}^{a} \rangle_{fi} ~,\
\label{e:divme}
\end{eqnarray}
{\it only} if the initial and final states $|\Psi_{f,i} \rangle$
are solutions to the nuclear Schr\" odinger equation 
$H|\Psi_{n} \rangle = E_{n}|\Psi_{n} \rangle$ (here 
$q_{0} = E_{f} - E_{i}$). We used the abbreviation
$\langle A \rangle_{fi} = \langle \Psi_{f}| A |\Psi_{i}\rangle$.
Thus, {\it the nuclear pion absorption/emission 
amplitude $\langle \Gamma_{\pi}^{a}\rangle_{fi}$ 
must be evaluated using a pion absorption/emission operator 
$\Gamma_{\pi}^{a}$ that matches the nuclear dynamics leading to
the nuclear wave functions  $|\Psi_{n} \rangle$, 
if the result is to agree with PCAC.} In the last line of Eq. 
(\ref{e:divme}) we used 
$q^{2} = q_{0}^{2} - {\bf q}^{2}$, which is true for elastic
scattering in the Breit frame, i.e., when 
$q_{0} = E_{f} - E_{i} = 0$.

The idea of consistency of nuclear wave functions and MECs is 
not a new one: it has been a part of the electronuclear physics 
``folklore" for some time. The
extension of this idea to the axial currents and pion production
amplitudes seems to be less well known. In the following we
shall apply this idea to nuclear axial currents in several
variations of the sigma model with $\omega$ mesons. 

\section{Axial currents in Sigma Models}

We shall use two of the sigma models already developed in 
Ref. \cite{ax96}. As is well known, the pseudoscalar and the
pseudovector, or gradient, $\pi NN$ couplings reduce to the same 
one-pion-exchange potential (OPEP) to leading order in 
nonrelativistic (NR) expansion. 
At first sight one might think that this implies identical 
axial MECs in the linear and nonlinear
sigma models. This is not so because the linear sigma model
OBEP includes a sigma-exchange potential as well, whereas the 
nonlinear one does not. 
Moreover, the difference between the ``heterotic sigma model" 
and the other two sigma models persists, although in an
unusual way: its (spatial) axial current is renormalized by a
factor $g_A$, though the axial charge is not.
In the following we look at each model separately. Subsequently 
we add the isoscalar-vector meson $\omega$, which plays an
important role by providing short-range repulsion and is 
chirally invariant by itself. That brings about new terms into 
the axial current and the pion production operator in a way that
is consistent with PCAC. 

\subsection{The one-body current}

The one-body part of the nuclear axial current is the sum over 
all nucleons of the {\it direct} axial current and the {\it the 
pion pole term}, see Fig. 3(a) in Ref. \cite{ax96}. In 
configuration space this takes its usual non-relativistic form 
\begin{eqnarray}
{\bf J}_{5}^{a}({\rm 1-body}) &=& g_A
\sum_{i=1}^{A} {\tau^{a}_{(i)} \over {2}} 
\left({\bbox{\sigma}}_{(i)} -  \bbox{\nabla}_{\bf R}
\left({{\bbox{\sigma}}_{(i)} \cdot \bbox{\nabla}_{\bf R}
\over{(\bbox{\nabla}_{\bf R}^{2} - m_{\pi}^2)
}}\right)\right) 
\delta({\bf R} - {\bf r}_{(i)})~~. \
\label{e:ax1b}
\end{eqnarray} 
The axial charge density 
\begin{eqnarray}
\rho^{a}_{5}({\rm 1-body}) &=& - i g_A 
\sum_{i=1}^{A} {\tau^{a}_{(i)} \over {2 M}} 
\left\{
\bbox{\sigma}_{(i)} \cdot {\bf \nabla}_{(i)}, 
\delta({\bf R} - {\bf r}_{(i)}) \right\} \
\label{e:axc1b}
\end{eqnarray}
on the other hand is taken {\it without} the pion-pole 
contribution. This simplifying 
assumption is justified {\it ex post facto} by the fact that it
does not prevent a successful construction of a consistent
approximation. 
This does not mean that one cannot find, or should not
search for approximations that avoid this assumption.

The axial coupling constant $g_{A}$ is either unity, as in the 
linear sigma model, or its measured value 1.26 
in the spatial part of the axial current Eq. (\ref{e:ax1b}), 
and unity in the axial charge Eq. (\ref{e:axc1b}) of the 
heterotic sigma model of Ref. \cite{ax96}. [For more on this, 
see Appendix \ref{app}.]
We also neglect all nucleon electroweak form factors. This should 
be adequate for the purpose of describing the static properties 
of the nucleus. The nuclear matrix element of the direct
one-body axial current is depicted in Fig. \ref{f:1'}(a),
whereas the pion-pole part can be constructed by attaching 
the axial current ``wavy line" to the external pion in Fig. 
\ref{f:1}(a).

\paragraph{Linear sigma model}
The $i^{th}$-nucleon current in the linear sigma model, 
{\it i.e.}, with a unit (``normalized") nucleon axial coupling,
in momentum space reads
\begin{eqnarray}
{\bf J}_{5, (i)}^{a}({\bf p}_{i}^{'}, {\bf p}_{i}) &=&  
{\tau_{(i)}^{a}\over 2}  \left[\bbox{\sigma}_{(i)} - 
{\bf q}
\left({\bbox{\sigma}_{(i)} \cdot {\bf q} \over{{\bf q}^{2} + m_{\pi}^2}} \right)
\right] \ ,
\label{e:1bl}
\end{eqnarray}
where ${\bf q}={\bf p}_{i}^{'}-{\bf p}_{i}$,
and satisfies the nonrelativistic (static) version of the 
single-fermion axial Ward-Takahashi identity (WT id.)
\begin{eqnarray}
{\bf q} \cdot {\bf J}_{5, (i)}{a}({\bf p}_{i}^{'}, {\bf p}_{i}) 
&=& 
\left({f_{\pi} m_{\pi}^2 \over{{\bf q}^{2} + m_{\pi}^2}} \right) 
g_{0} \tau_{(i)}^{a}
\left({\bbox{\sigma}_{(i)} \cdot {\bf q} \over 2M}\right)
\nonumber \\
&=& 
i \left({f_{\pi} m_{\pi}^2 \over{q^{2} - m_{\pi}^2}} \right) 
\Gamma_{\pi}^{a}({\bf p}_{i}^{'}, {\bf p}_{i}; {\rm 1-body})
\nonumber \\
&\simeq& 
- i 
\left({f_{\pi} m_{\pi}^2 \over{{\bf q}^{2} + m_{\pi}^2}}\right) 
\Gamma_{\pi}^{a}({\bf p}_{i}^{'}, {\bf p}_{i}; {\rm 1-body})
~,\ 
\label{e:wi1bl}
\end{eqnarray}
which follows from the Goldberger-Treiman (GT) relation 
$M = g_{0} f_{\pi}$. We use the same symbol for operators in
configuration and momentum space.
The second line on the right hand-side (r.h.s.) of Eq. 
(\ref{e:wi1bl}) is the single-nucleon pion absorption operator
multiplied by the divergence of the axial current factor 
$f_{\pi} m_{\pi}^2$
and the static pion propagator $({\bf q}^{2} + m_{\pi}^{2})^{-1}$.
We see that the pion absorption operator arises naturally from the 
divergence of the axial current. Both in this and in the 
heterotic sigma model the one-body axial charge
operator Eq. (\ref{e:axc1b}) does not contain the factor $g_A$,
i.e., in momentum space it reads 
\begin{eqnarray}
{\rho}_{5, (i)}^{a}({\bf p}_{i}^{'}, {\bf p}_{i}) &=&  
{\tau_{(i)}^{a}\over 2} 
\bbox{\sigma}_{(i)} \cdot 
\left({{\bf p}_{i}^{'} + {\bf p}_{i} \over{2M}} \right)
\ ,
\label{e:ac1bl}
\end{eqnarray}
for proof see Appendix \ref{app}.

\paragraph{Heterotic sigma model}
In the heterotic sigma model the
GT relation is modified to $g_A M = g_{\pi NN} f_{\pi}$, where 
$g_A = 1.26$ and the one-body current becomes
\begin{eqnarray}
{\bf J}_{5, (i)}^{a}({\bf p}_{i}^{'}, {\bf p}_{i}) 
&=& 
g_A  {\tau_{(i)}^{a}\over 2}  \left[\bbox{\sigma}_{(i)} 
- {\bf q} \left({\bbox{\sigma}_{(i)} \cdot {\bf q} 
\over{{\bf q}^{2} + m_{\pi}^2}} \right)
\right] \ ,
\label{e:1bnl}
\end{eqnarray}
which satisfies the single-nucleon axial Ward-Takahashi 
identity (WT id.)
\begin{eqnarray}
{\bf q} \cdot {\bf J}_{5, (i)}^{a}({\bf p}_{i}^{'}, {\bf p}_{i}) 
&=& 
f_{\pi} \left({m_{\pi}^2 \over{{\bf q}^{2} + m_{\pi}^2}} 
\right) g_{\pi NN} \tau_{(i)}^{a}
\left({\bbox{\sigma}_{(i)} \cdot {\bf q}\over 2M}\right) 
\nonumber \\
&=& 
i \left({f_{\pi} m_{\pi}^2 \over{q^{2} - m_{\pi}^2}} \right)
\Gamma_{\pi}^{a}({\bf p}_{i}^{'}, {\bf p}_{i}; {\rm 1-body})
\nonumber \\
&\simeq& 
- i 
\left({f_{\pi} m_{\pi}^2 \over{{\bf q}^{2} + m_{\pi}^2}}\right)
\Gamma_{\pi}^{a}({\bf p}_{i}^{'}, {\bf p}_{i}; {\rm 1-body})
~,\ 
\label{e:wi1bnl}
\end{eqnarray}
since 
$\left[~T,~\rho^{a}_{5} ({\rm 1-body})\right] = 0 + 
{\cal O}(M^{-2})$.
The spatial parts of one-nucleon axial currents in the two models 
are almost identical in the NR limit, the only difference being 
the overall factor $g_A$. As stated above, the axial charges 
in the two models are identical, (see Appendix \ref{app}).
  
\subsection{Two-nucleon axial current}

Having shown that in order to have a partially conserved 
axial current in a non-relativistic nuclear model, 
one must have two-nucleon axial currents, we turn towards
constructing such MECs. The two-body currents 
${\bf J}_{5}({\rm 2-b})$ 
appropriate to the meson exchange two-nucleon potentials will be 
constructed and we shall show that they lead to 
PCAC. We construct these 
``meson exchange currents" by non-relativistic reduction of 
covariant Feynman amplitudes in specific chiral models. 

Due to the presence of gradient operators it will 
be to our advantage to work in the momentum space.
Definition of the Fourier transform of two-body currents 
into momentum space 
can be found in Ref. \cite{oht79}, among others. 
The current conservation relation in momentum space can now be 
written as
\begin{eqnarray} 
{\bf q} \cdot {\bf J}_{5}^{a}({\rm 2-body})
&=&
\left[V({\rm 2-body}),~\rho_{5}^{a}({\rm 1-body}) \right] 
- i 
\left({f_{\pi} m_{\pi}^2 \over{{\bf q}^{2} + m_{\pi}^2}} \right)
\Gamma_{\pi}^{a}({\rm 2-body})~.\ 
\label{e:div2bm}
\end{eqnarray}
We repeat that this equation follows from only two
assumptions: (i) PCAC, and (ii) quantum mechanics.
As stated above, the only ``sure-fire" way of constructing an 
axial MEC that satisfies PCAC that we know of is to start from a 
relativistic chiral Lagrangian model. 
We shall use the two variations of the linear sigma model 
already utilized in Ref. \cite{ax96} and the simplest 
$\omega NN$ interaction Lagrangian that preserves chiral
symmetry.

\subsubsection{Linear sigma model}

To construct the partially conserved nonrelativistic axial 
two-nucleon current in this model
we start from the corresponding covariant amplitude.
A nonrelativistic expansion in powers of $1/M$ leads to three
terms of $O(1/M)$: (i) one due to the meson-in-flight diagrams, 
Fig. (\ref{f:2'}) plus their pion-pole counterparts, Fig. 
(\ref{f:2}), (the covariant current that defines this amplitude 
can be found in Eqs. (33), (34) of Ref. \cite{ax96}):
\begin{eqnarray}
{\bf A}_{\rm I}^{a} ({\bf k}_{1}, {\bf k}_{2}, {\bf q}) 
&=& 
{g_{0}^{2} \over 2 M}
\tau_{(1)}^{a} 
\biggl[{\bf k}_{1} - {\bf k}_{2} + {\bf q} 
\left({f_{\pi} g_{\sigma \pi \pi} 
\over{{\bf q}^{2} + m_{\pi}^2}}\right) \biggr]
\nonumber \\
&\times& 
{\bbox{\sigma}_{(1)} \cdot {\bf k}_{1}
\over{\left({\bf k}_{2}^{2} + m_{\sigma}^2\right) 
\left({\bf k}_{1}^{2} + m_{\pi}^2\right)}}
+  \left(1 \leftrightarrow 2 \right) 
~, \
\label{e:mec1}
\end{eqnarray}
where ${\bf k}_{i} = {\bf p}_{i} - {\bf p}_{i}^{'}$, $i = 1, 2$;
and $f_{\pi} g_{\sigma \pi \pi} = m_{\sigma}^2 - m_{\pi}^2$.
Three-momentum conservation reads ${\bf k}_{1} + {\bf k}_{2}
+ {\bf q} = 0$.
This MEC alone is not sufficient for PCAC, as can be seen from 
the corresponding divergence,
 which reads
\begin{eqnarray} 
{\bf q} \cdot {\bf A}_{\rm I}^{a} ({\bf k}_{1}, {\bf k}_{2}, 
{\bf q})
&=& 
{g_{0}^{2} \over 2 M} \tau_{(1)}^{a} 
\biggl[
{\bf k}_{2}^{2} - {\bf k}_{1}^{2} + 
\left(m_{\sigma}^2 - m_{\pi}^2\right) - f_{\pi} m_{\pi}^2
\left({g_{\sigma \pi \pi} \over{{\bf q}^{2} + m_{\pi}^2}}
\right) \biggr]
\nonumber \\
&\times& 
{\bbox{\sigma}_{(1)} \cdot {\bf k}_{1} \over{
\left({\bf k}_{2}^{2} + m_{\sigma}^2\right) 
\left({\bf k}_{1}^{2} + m_{\pi}^2\right)}}
+ \left(1 \leftrightarrow 2 \right)
\nonumber \\
&=& 
- {g_{0}^{2} \over 2M} \left[
\tau_{(1)}^{a} {\bbox{\sigma}_{(1)} \cdot {\bf k}_{1} 
\over{{\bf k}_{2}^{2} + m_{\sigma}^2}}
+ \left(1 \leftrightarrow 2 \right)\right] 
\nonumber \\
&+& 
{g_{0}^{2} \over 2M} \left[
\tau_{(1)}^{a} {\bbox{\sigma}_{(1)} \cdot {\bf k}_{1} 
\over{{\bf k}_{1}^{2} + m_{\pi}^2}}
+ \left(1 \leftrightarrow 2 \right)\right] 
+ {\cal O}(f_{\pi}m_{\pi}^2)
\nonumber \\
&=& 
\left[V_{\sigma},~\rho_{5}^{a} \right]  
\nonumber \\
&+& 
{g_{0}^{2} \over 2M} \left[\tau_{(1)}^{a} 
{\bbox{\sigma}_{(1)} \cdot {\bf k}_{1} 
\over{{\bf k}_{1}^{2} + m_{\pi}^2}}
+ \left(1 \leftrightarrow 2 \right)\right] 
\nonumber \\ 
&+& 
 {g_{0}^{2} \over 2M} \left[ \tau_{(1)}^{a} 
{\bbox{\sigma}_{(1)} \cdot {\bf q} 
\over{{\bf k}_{2}^{2} + m_{\sigma}^2}}  
+ \left(1 \leftrightarrow 2 \right)\right]
\nonumber \\
&+& {\cal O}(f_{\pi}m_{\pi}^2)~,\
\label{e:comm} 
\end{eqnarray}
where the one-meson-exchange $NN$ potential in the linear sigma
model reads
\begin{eqnarray} 
V_{\rm 2-b}\left({\bf p}\right)
&=& 
V_{\sigma}\left({\bf p}\right)  +
V_{\pi}\left({\bf p}\right)  
\nonumber \\
&=&-
{g_{0}^{2} \over{{\bf p}^{2} + m_{\sigma}^2}}
\nonumber \\
&+& 
{\vec \tau}_{(1)} \cdot {\vec \tau}_{(2)}  
\left({{\bbox{\sigma}_{(1)}} \cdot {\bf p} \over 2 M}\right) 
\left({{\bbox{\sigma}_{(2)}} \cdot {\bf p} \over 2 M}\right) 
\left({g_{0}^{2} \over{{\bf p}^{2} + m_{\pi}^2}}\right) 
\nonumber \\
&=& - 
{g_{0}^{2} \over{{\bf p}^{2} + m_{\sigma}^2}}
+ {\cal O}(1/M^2)
~. \
\label{e:pots}
\end{eqnarray}
Note that the $\sigma$-exchange potential is of $O(1/M^{0} = 1)$, 
whereas the $\pi$-exchange potential is of $O(1/M^2)$.
Consequently, the commutator of the potential and the 
one-body axial charge also fall into two distinct orders
in $1/M$. This fact allows an apparently clean and simple 
separation of the $\sigma$-exchange current effects 
from the $\pi$-exchange ones.
In this paper we are primarily interested in the
$\sigma$-exchange currents, so we leave the $\pi$-induced
ones aside, as they have been extensively 
studied in the literature \cite{oht79,sauer92}.

The commutator on the right-hand side of Eq. (\ref{e:div2bm}) is 
\begin{eqnarray} 
\left[V_{\rm 2-b},~\rho_{5}^{a} \right]
&=& 
\left[V_{\sigma},~\rho_{5}^{a} \right] 
+ {\cal O}(1/M^3)
\nonumber \\ 
&=& 
{g_{0}^{2} \over 2 M} \left[\tau_{(1)}^{a} 
{\bbox{\sigma}_{(1)} \cdot {\bf k}_{2} 
\over{{\bf k}_{2}^{2} + m_{\sigma}^2}}
+ \left(1 \leftrightarrow 2 \right)\right]
+ {\cal O}(1/M^3) ~. \
\label{e:lscom}
\end{eqnarray}
We see that 
the $\sigma$-exchange leads to terms of $O(1/M)$, whereas the 
$\pi$-exchange leads to terms of $O(1/M^3)$ in the commutator.
Comparing Eq. (\ref{e:lscom}) with the divergence of the axial 
two-body current Eq. (\ref{e:comm}) we see that we are rather 
far from having the sigma-exchange potential commutator on the 
right-hand side. There are four terms left over:
two are proportional to the $\pi$ propagator thus indicating
perhaps a relationship to the $\pi$-exchange current, but also
being of ${\cal O}(1/M)$, i.e., two orders in $1/M$ lower
than the lowest expected $\pi$-exchange current contribution! 
The other two are proportional to the $\sigma$ propagator. In
other words, our expectations do not square well with these 
initial results. 
So, the question is: whence do these ``extra" terms come from 
and is there something that might compensate for them? 
The answer is that among the apparently ``higher-order" 
terms in the axial $\pi$-exchange currents there are two such
ones.
 
(i) One is the $\pi$-exchange axial current, Fig. \ref{f:1}(b),
\begin{eqnarray}
{\bf A}_{\rm II}^{a} ({\bf k}_{1}, {\bf k}_{2}, {\bf q}) 
&=& 
- {g_{0}^{3} \over 2 M^2}  
\left({f_{\pi} {\bf q} \over{{\bf q}^{2} + m_{\pi}^2}} \right)
\left[\tau_{(1)}^{a} {\bbox{\sigma}_{(1)} \cdot {\bf k}_{1}
\over{{\bf k}_{1}^{2} + m_{\pi}^2}}
+ \left(1 \leftrightarrow 2 \right) \right]
~, \
\label{e:mec2}
\end{eqnarray}
that is actually one order in $1/M$ lower than naively expected,
due to the validity of the Goldberger-Treiman relation
$f_{\pi} g_{0} = M$ in the linear sigma model. Thus 
\begin{eqnarray}
{\bf A}_{\rm II}^{a} ({\bf k}_{1}, {\bf k}_{2}, {\bf q}) 
&=& 
- {g_{0}^{2} \over 2 M}  
\left({{\bf q} \over{{\bf q}^{2} + m_{\pi}^2}} \right)
\left[\tau_{(1)}^{a} {\bbox{\sigma}_{(1)} \cdot {\bf k}_{1}
\over{{\bf k}_{1}^{2} + m_{\pi}^2}}
+ \left(1 \leftrightarrow 2 \right) \right] 
~. \
\label{e:mec2'}
\end{eqnarray}
This current is due to an effective two-pion-nucleon vertex, 
which is really a time-ordered Z-graph arising in the 
NR reduction of the nucleon Feynman propagator (the so-called
pair current"), 
and is not an ``elementary" interaction term in the Lagrangian.
The divergence of this axial MEC is 
\begin{eqnarray}
{\bf q} \cdot {\bf A}_{\rm II}^{a} ({\bf k}_{1}, {\bf k}_{2}, 
{\bf q})
&=& 
- {g_{0}^{2} \over 2 M} \left[ \tau_{(1)}^{a}   
{\bbox{\sigma}_{(1)} \cdot {\bf k}_{1}
\over{{\bf k}_{1}^{2} + m_{\pi}^2}}
+ \left(1 \leftrightarrow 2 \right) \right]
+ {\cal O}(f_{\pi} m_{\pi}^2)~. \
\label{e:mec2d}
\end{eqnarray}

(ii) Similarly, there is a $\sigma$-exchange ``Z-graph", Fig. 
\ref{f:3}, as well
\begin{eqnarray}
{\bf A}_{\rm III}^{a} ({\bf k}_{1}, {\bf k}_{2}, {\bf q}) 
&=& 
- {g_{0}^{2} \over 2M} 
\left({{\bf q} \over{{\bf q}^{2} + m_{\pi}^2}} \right)
\left[\tau_{(1)}^{a} {\bbox{\sigma}_{(1)} \cdot {\bf q} 
\over{{\bf k}_{2}^{2} + m_{\sigma}^2}} + 
\left(1 \leftrightarrow 2 \right) \right]~. \
\label{e:mec3}
\end{eqnarray}
Its divergence is 
\begin{eqnarray}
{\bf q} \cdot {\bf A}_{\rm III}^{a}({\bf k}_{1}, 
{\bf k}_{2}, {\bf q}) 
&=& 
- {g_{0}^{2} \over 2M} \left[ \tau_{(1)}^{a} 
{\bbox{\sigma}_{(1)} \cdot {\bf q} 
\over{\left({\bf k}_{2}^{2} + m_{\sigma}^2\right)}}  
+ \left(1 \leftrightarrow 2 \right)\right]
+ {\cal O}(f_{\pi}m_{\pi}^2)~, \
\label{e:mec3d}
\end{eqnarray}
which is exactly what we need to bring the
continuity equation (\ref{e:comm}) in the linear sigma model
into the canonical form (\ref{e:div2bm}).

Thus, the complete nonrelativistic axial MEC to ${\cal O}(1/M)$ 
in the linear sigma model is
\begin{eqnarray}
{\bf J}_{2-b}^{a}({\bf k}_{1}, {\bf k}_{2}, {\bf q})
&=& 
{\bf A}_{\rm I}^{a}({\bf k}_{1}, {\bf k}_{2}, {\bf q}) +
{\bf A}_{\rm II}^{a}({\bf k}_{1}, {\bf k}_{2}, {\bf q}) +
{\bf A}_{\rm III}^{a}({\bf k}_{1}, {\bf k}_{2}, {\bf q}) 
\nonumber \\
&=&
{g_{0}^{2} \over 2 M}
\tau_{(1)}^{a} 
\Bigg\{
\left[{\bf k}_{1} - {\bf k}_{2} + {\bf q} 
\left({f_{\pi} g_{\sigma \pi \pi} 
\over{{\bf q}^{2} + m_{\pi}^2}}\right) \right]
\nonumber \\
&\times& 
{\bbox{\sigma}_{(1)} \cdot {\bf k}_{1}
\over{\left({\bf k}_{2}^{2} + m_{\sigma}^2\right) 
\left({\bf k}_{1}^{2} + m_{\pi}^2\right)}}
\nonumber \\
&-& 
\left({{\bf q} \over{{\bf q}^{2} + m_{\pi}^2}} \right)
\left[{\bbox{\sigma}_{(1)} \cdot {\bf k}_{1}
\over{{\bf k}_{1}^{2} + m_{\pi}^2}}
+ {\bbox{\sigma}_{(1)} \cdot {\bf q} 
\over{{\bf k}_{2}^{2} + m_{\sigma}^2}} \right] \Bigg\}
+ \left(1 \leftrightarrow 2 \right)
~. \
\label{e:lsmec}
\end{eqnarray}
Thus, Eq. (\ref{e:lsmec}) uniquely fixes the pion 
production/absorption operator $\Gamma_{\pi}^a$ in the 
one-boson-exchange
potential approximation to the linear sigma model. 
This operator is to be used together with the 
${\cal O}(1/M^3)$ $\pi$-exchange operators and 
nuclear wave functions that are solutions to the Schr\" odinger
equation with the above one pion + sigma exchange NN potential. 

The resulting MEC operator is perhaps somewhat unexpected: 
certainly 
the first (``meson-in-flight") term is not a surprise, but the 
presence of the second (``pion seagull") term might seem a 
little odd at first: 
One would have been hard pressed to correctly guess the second term 
in the MEC without the benefit of guidance by the linear 
sigma model.
Thus we have expanded the nuclear interaction to include
one-sigma-exchange NN potential (beside the OPEP) in a manner that
is consistent with PCAC, but still with $g_A = 1$. Next, we shall
relax that assumption. Once again, we resort to a chiral
Lagrangian model for guidance.

\subsubsection{Heterotic sigma model}

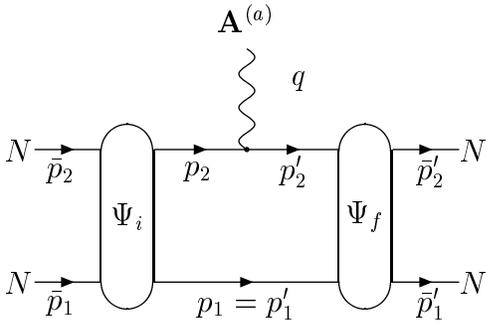
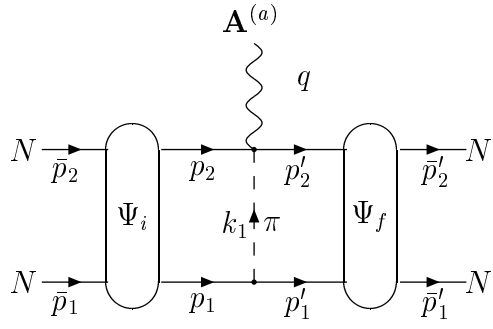
\begin{figure}
\caption{
Effective nonrelativistic 
Feynman diagrams contributing to the one- (a) and the two-body 
axial current nuclear matrix element (b). Each graph consists
of a ``direct" term and a ``pion pole" term. We display
only the direct terms here.
In the nonlinear sigma model there are only graphs 1(a) and 1(b).
The dashed line denotes a pion, the solid one a nucleon, the
wavy line is the external axial current (source).
\label{f:1'}}
\end{figure}

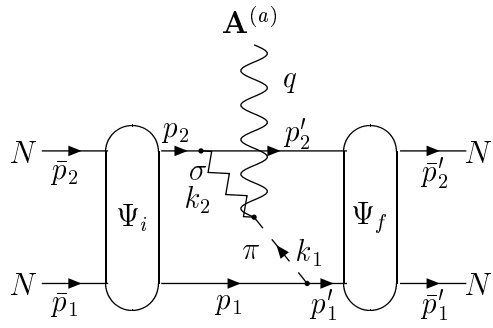
\begin{figure}
\caption{The ``meson-in-flight" graph contributing to the
axial current in the 
linear and the heterotic sigma models. The zig-zag line 
denotes a sigma meson. One must keep all four graphs in Figs. 
1(a),(b) and 2, 3 in both the linear and the heterotic sigma 
model. 
\label{f:2'}}
\end{figure}

The heterotic sigma model differs from the linear one by the 
presence of $g_A$ in the GT relation, 
$g_{\pi NN} f_{\pi} = g_{A} M$, and in the spatial part of
the axial current (\ref{e:ax1b}), but {\it not} in the axial
charge (\ref{e:axc1b}). [For a proof of this statement, see 
Appendix \ref{app}.] This variation induces some curious changes
in the axial two-nucleon currents. 
To begin with, there are two new types 
of relativistic Born approximation Feynman diagrams, depicted
in Figs. \ref{f:1'}(b), \ref{f:1}(b), and Fig. \ref{f:3}. 
[The complete relativistic current
consists of the sum of Eqs. (41), (47) and (51), and its 
divergence in the sum of Eqs. (42), (48) and (52) in Ref.
\cite{ax96}.] 
By evaluating its
matrix element between on-shell single-nucleon states and making
the nonrelativistic expansion in powers of $1/M$, one finds a 
host of $\pi$-exchange currents of $O(1/M^3)$, and two different
$\sigma$-exchange currents of $O(1/M)$. One of them is the 
familiar meson-in-flight diagram (+ its exchange), Fig.
\ref{f:2'}, but rescaled by a factor of $g_A$:
\begin{eqnarray}
{\bf A}_{\rm I}^{a}({\bf k}_{1}, {\bf k}_{2}, {\bf q}; h) 
&=& 
g_{A} {g_{0}^{2} \over 2 M} \tau_{(1)}^{a} 
\biggl[{\bf k}_{1} - {\bf k}_{2} + {\bf q} 
\left({f_{\pi} g_{\sigma \pi \pi} 
\over{{\bf q}^{2} + m_{\pi}^2}}\right) \biggr]
\nonumber \\
&\times& 
{\bbox{\sigma}_{(1)} \cdot {\bf k}_{1}
\over{\left({\bf k}_{2}^{2} + m_{\sigma}^2\right) 
\left({\bf k}_{1}^{2} + m_{\pi}^2\right)}}
+ \left(1 \leftrightarrow 2 \right)~, \
\label{e:hmec1}
\end{eqnarray}
for which 
the current divergence in momentum space reads
\begin{eqnarray} 
{\bf q} \cdot {\bf A}_{\rm I}^{a}({\bf k}_{1}, {\bf k}_{2}, 
{\bf q}; h)
&=& 
g_{A} {g_{0}^{2} \over 2 M} \tau_{(1)}^{a} 
\biggl[{\bf k}_{2}^{2} - {\bf k}_{1}^{2} + {\bf q}^{2} 
\left({f_{\pi} g_{\sigma \pi \pi} 
\over{{\bf q}^{2} + m_{\pi}^2}}\right) \biggr]
\nonumber \\
&\times& 
{\bbox{\sigma}_{(1)} \cdot {\bf k}_{1}
\over{\left({\bf k}_{2}^{2} + m_{\sigma}^2\right) 
\left({\bf k}_{1}^{2} + m_{\pi}^2\right)}}
+ \left(1 \leftrightarrow 2 \right)
\nonumber \\   
&=& 
g_{A} \left[V_{\sigma},~\rho_{5}^{a} \right] 
\nonumber \\
&+& 
g_{A} {g_{0}^{2} \over 2M} \left[\tau_{(1)}^{a} 
{\bbox{\sigma}_{(1)} \cdot {\bf k}_{1} 
\over{{\bf k}_{1}^{2} + m_{\pi}^2}}
+ \left(1 \leftrightarrow 2 \right)\right] 
\nonumber \\ 
&+& 
g_{A} {g_{0}^{2} \over 2M} \left[ \tau_{(1)}^{a} 
{\bbox{\sigma}_{(1)} \cdot {\bf q} 
\over{{\bf k}_{2}^{2} + m_{\sigma}^2}}  
+ \left(1 \leftrightarrow 2 \right)\right]
\nonumber \\
&+&   {\cal O}(f_{\pi} m_{\pi}^2)~.\
\label{e:hcomm} 
\end{eqnarray}
The one-meson-exchange $NN$ potential in the heterotic 
sigma model reads
\begin{eqnarray} 
V_{\rm 2-b}\left({\bf p}\right)
&=& 
V_{\sigma}\left({\bf p}\right)  +
V_{\pi}\left({\bf p}\right)  
\nonumber \\
&=&-
{g_{0}^{2} \over{{\bf p}^{2} + m_{\sigma}^2}}
\nonumber \\
&+& 
{\vec \tau}_{(1)} \cdot {\vec \tau}_{(2)}  
\left({{\bbox{\sigma}_{(1)}} \cdot {\bf p} \over 2 M}\right) 
\left({{\bbox{\sigma}_{(2)}} \cdot {\bf p} \over 2 M}\right) 
\left({g_{\pi NN}^{2} \over{{\bf p}^{2} + m_{\pi}^2}}\right) 
\nonumber \\
&=& - 
{g_{0}^{2} \over{{\bf p}^{2} + m_{\sigma}^2}}
+ {\cal O}(1/M^2) ~, \
\label{e:hpot}
\end{eqnarray}
and $g_{\pi NN} = g_{A} g_{0}$, due to the new GT relation.

Similarly, the pion ``Z-graph", Fig. \ref{f:1'}(b), contribution 
is also renormalized by factor $g_A$:
\begin{eqnarray}
{\bf A}_{\rm II}^{a}({\bf k}_{1}, {\bf k}_{2}, {\bf q};h) 
&=& 
- g_{A} {g_{0}^{3} \over 2 M^2}  
\left({{\bf q} f_{\pi} \over{{\bf q}^{2} + m_{\pi}^2}} \right)
\left[\tau_{(1)}^{a} {\bbox{\sigma}_{(1)} \cdot {\bf k}_{1}
\over{{\bf k}_{1}^{2} + m_{\pi}^2}}
+ \left(1 \leftrightarrow 2 \right) \right]
\nonumber \\
&=& 
- g_{A} {g_{0}^{2} \over 2 M}  
\left({{\bf q} \over{{\bf q}^{2} + m_{\pi}^2}} \right)
\left[\tau_{(1)}^{a} {\bbox{\sigma}_{(1)} \cdot {\bf k}_{1}
\over{{\bf k}_{1}^{2} + m_{\pi}^2}}
+ \left(1 \leftrightarrow 2 \right)\right]~. \
\label{e:hmec2}
\end{eqnarray}
The divergence of this axial MEC is 
\begin{eqnarray}
{\bf q} \cdot {\bf A}_{\rm II}^{a} ({\bf k}_{1}, {\bf k}_{2}, 
{\bf q}; h)
&=& 
- g_{A} {g_{0}^{2} \over 2 M} \left[ \tau_{(1)}^{a}   
{\bbox{\sigma}_{(1)} \cdot {\bf k}_{1}
\over{{\bf k}_{1}^{2} + m_{\pi}^2}}
+ \left(1 \leftrightarrow 2 \right) \right]
+ {\cal O}(f_{\pi} m_{\pi}^2)~. \
\label{e:hmec2d}
\end{eqnarray}

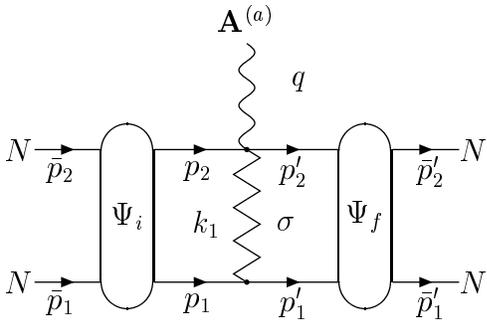
\begin{figure}
\caption{ 
Sigma-meson-exchange current contributing to 
the nuclear pion production matrix element. It consists of a
``pair", or Z-graph time-ordered contribution in the linear
sigma model to which the elementary $\pi \sigma NN$ vertex is
added in the heterotic model.
\label{f:3'}}
\end{figure}

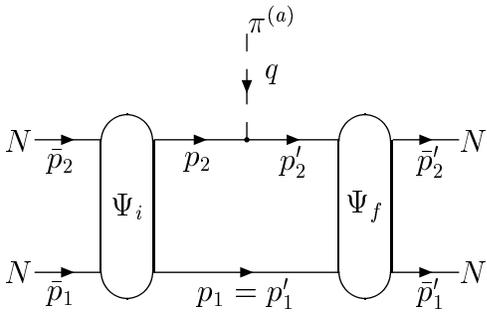
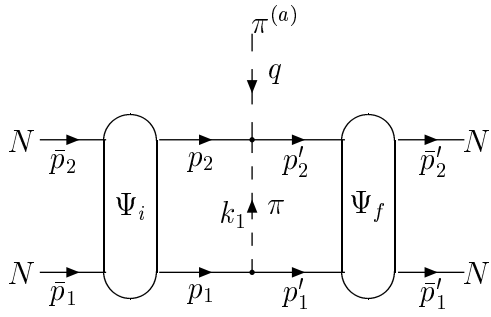
\begin{figure}

\caption{ Effective nonrelativistic 
Feynman diagrams contributing to the one- (a) and the two-body 
pion production nuclear matrix element (b). 
In the nonlinear sigma model there are only graphs 1(a) and 1(b).
The pion-exchange current (b) consists of a
``pair", or Z-graph time-ordered contribution in the linear
sigma model to which the elementary $\pi \pi NN$ vertex is
added in the heterotic model.
The dashed line denotes a pion, the solid one a nucleon.
\label{f:1}}
\end{figure}

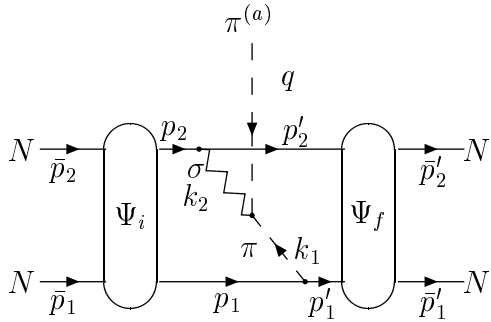
\begin{figure}
\caption{The ``meson-in-flight" graph contributing to the
linear and the heterotic sigma models. The zig-zag line 
denotes a sigma meson. One must keep all four graphs in Figs. 
1(a),(b) and 2, 3 in both the linear and the heterotic sigma 
model. 
\label{f:2}}
\end{figure}

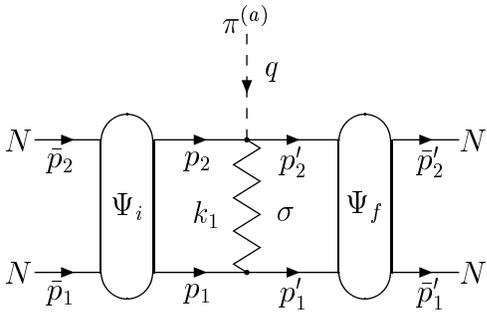
\begin{figure}
\caption{ 
Sigma-meson-exchange current contributing to 
the nuclear pion production matrix element. It consists of a
``pair", or Z-graph time-ordered contribution in the linear
sigma model to which the elementary $\pi \sigma NN$ vertex is
added in the heterotic model.
\label{f:3}}
\end{figure}

Finally, in the heterotic sigma model in addition to the 
$\sigma$-exchange ``Z-graph", Fig. \ref{f:3}, 
contribution 
\begin{eqnarray}
{\bf A}_{\rm III}^{a} ({\bf k}_{1}, {\bf k}_{2}, 
{\bf q}; h) 
&=& 
- {g_{0}^{2} \over 2M} \tau_{(1)}^{a} 
\left({{\bf q} \over{{\bf q}^{2} + m_{\pi}^2}} \right)
{\bbox{\sigma}_{(1)} \cdot {\bf q} 
\over{{\bf k}_{2}^{2} + m_{\sigma}^2}} + 
\left(1 \leftrightarrow 2 \right)~, \
\label{e:hmec3}
\end{eqnarray}
there is also an elementary $A_{\mu}^{a} \sigma NN$ axial
current vertex,  Fig. \ref{f:3'}, and a $\pi \sigma NN$
vertex-induced MEC, Fig. \ref{f:3},
\begin{eqnarray}
{\bf A}_{\rm IV}^{a} ({\bf k}_{1}, {\bf p}_{2}, 
{\bf q}; h) 
&=& 
- f_{\pi} \left({g_{A} - 1 \over f_{\pi}}\right)
{g_{0}^2 \over 2M}  {\tau_{(1)}^{a}
\over{\left({\bf k}_{2}^{2} + m_{\sigma}^2 \right)}} 
\nonumber \\
&\times&  
\left[2 \bbox{\sigma}_{(1)} + 
\left({{\bf q} \over{{\bf q}^{2} + m_{\pi}^2}} \right)
\bbox{\sigma}_{(1)} \cdot \left({\bf k}_{2} - {\bf q}\right) 
\right] + 
\left(1 \leftrightarrow 2 \right)~. \
\label{e:hmec4}
\end{eqnarray}
The divergence of the sum of these last two axial MECs is 
\begin{eqnarray}
{\bf q} \cdot {\bf A}_{\rm III + IV}^{a} 
({\bf k}_{1}, {\bf k}_{2}, {\bf q};h) 
&=& 
- {g_{0}^2 \over{2M}} 
{\tau_{(1)}^{a} \over{{\bf k}_{2}^{2} + m_{\sigma}^2}}
\Bigg[
(g_{A} - 1) \bbox{\sigma}_{(1)} \cdot 
\left({\bf k}_{2} + {\bf q}\right) 
+ \bbox{\sigma}_{(1)} \cdot {\bf q} \Bigg]
\nonumber \\
&+& 
\left(1 \leftrightarrow 2 \right)
+ {\cal O}(f_{\pi}m_{\pi}^2)~,
\label{e:dhmec3} \
\end{eqnarray}
which is exactly what we need to complete the continuity
equation (\ref{e:div2bm}).

The sum of these three axial MECs together with the one-body
(impulse approximation) terms plus the ${\cal O}(1/M^2)$ 
pion-exchange currents which we consistently suppressed in this 
paper, constitute the complete, PCAC-obeying axial current: 
\begin{eqnarray}
{\bf J}_{5, \rm 2-b}^{a} ({\bf k}_{1}, {\bf k}_{2}, {\bf q}; h)
&=& 
{\bf A}_{\rm I}^{a} + {\bf A}_{\rm II}^{a} +
{\bf A}_{\rm III}^{a} + {\bf A}_{\rm IV}^{a}
\nonumber \\
&=&
{g_{0}^{2} \over 2 M}
\tau_{(1)}^{a} 
\Bigg\{g_{A}
\left[{\bf k}_{1} - {\bf k}_{2} + {\bf q} 
\left({f_{\pi} g_{\sigma \pi \pi} 
\over{{\bf q}^{2} + m_{\pi}^2}}\right) \right]
\nonumber \\
&\times& 
{\bbox{\sigma}_{(1)} \cdot {\bf k}_{1}
\over{\left({\bf k}_{2}^{2} + m_{\sigma}^2\right) 
\left({\bf k}_{1}^{2} + m_{\pi}^2\right)}}
\nonumber \\
&-& 
\left({{\bf q} \over{{\bf q}^{2} + m_{\pi}^2}} \right)
\left[g_{A}{\bbox{\sigma}_{(1)} \cdot {\bf k}_{1}
\over{{\bf k}_{1}^{2} + m_{\pi}^2}}
+ {\bbox{\sigma}_{(1)} \cdot {\bf q} 
\over{{\bf k}_{2}^{2} + m_{\sigma}^2}} \right]
\nonumber \\
&-& 
\left({{g_{A} - 1} \over{{\bf k}_{2}^{2} + m_{\sigma}^2 }}\right) 
\left[2 \bbox{\sigma}_{(1)} + 
\left({{\bf q} \over{{\bf q}^{2} + m_{\pi}^2}} \right)
\bbox{\sigma}_{(1)} \cdot \left({\bf k}_{2} - {\bf q}\right) 
\right]  \Bigg\}
\nonumber \\
&+& 
\left(1 \leftrightarrow 2 \right)
~. \
\label{e:hsmec}
\end{eqnarray}
Most of this current is new: specifically, terms (II, III, IV)
have not been considered before, only the first term (I) having
been derived in Refs. \cite{tr94,hor94}.
This current satisfies PCAC
\begin{eqnarray}
{\bf q} \cdot 
{\bf J}_{5, \rm 2-b}^{a}({\bf k}_{1}, {\bf k}_{2}, {\bf q}; h) 
&=&
\left[V_{\sigma},~\rho_{5}^{a} \right] 
+   {\cal O}(f_{\pi}m_{\pi}^2)~.\
\label{e:hsdiv}
\end{eqnarray}
{\it This equation uniquely fixes the pion production/absorption
operator} in the OBEP approximation to the nuclear dynamics in 
the heterotic sigma model as the ``${\cal O}(f_{\pi}m_{\pi}^2)$" 
term, to be evaluated in the next section. 
We see that, by a curious turn of events, the divergence of 
the complete axial current (\ref{e:hsdiv})
equals the divergence found in the linear sigma model 
(\ref{e:hcomm}), 
modulo different ${\cal O}(f_{\pi}m_{\pi}^2)$ terms, of course.
In other words, we have exactly the same commutator of the 
one-sigma-exchange potential (O$\Sigma$EP) and the axial charge 
in these two sigma models, despite manifest differences between 
their axial currents. 
This fact is a consequence of identical axial charges and 
O$\Sigma$EP in the two models, the former being a subtle effect 
explained in \ref{app}. 
Finally we turn to the $\omega$ exchange.

\subsubsection{Omega meson exchange}

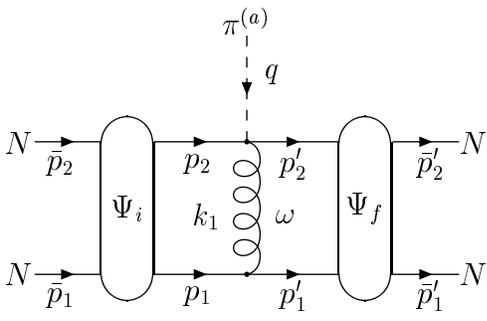
\begin{figure}
\caption{ 
The $\omega$-exchange current contributing to 
the nuclear pion production matrix element. 
The curly line denotes an omega meson. This graph  
can be added to all sigma models without disturbing their 
chiral symmetry. 
\label{f:4}}
\end{figure}

One may include the $\omega$-meson-exchange potential 
\begin{eqnarray} 
V_{\omega}\left({\bf p}\right) 
&=&
{g_{\omega}^{2} \over{{\bf p}^{2} + m_{\omega}^2}}
\label{e:omegapot} \
\end{eqnarray}
into the nuclear Hamiltonian $H$ at no peril 
to chiral symmetry because it is an isoscalar vector field which
is chirally invariant by itself in both the linear and the
nonlinear realization. As is well known, the main benefit of 
including this term into the nucleon-nucleon potential is that 
it provides short range repulsion that is otherwise absent from
the sigma models. The nuclear potential with $\pi, \sigma,\omega$ 
exchange has sufficient attraction to bind the deuteron
and enough repulsion to keep it weakly bound.
The associated axial MEC is, Fig. \ref{f:4},
\begin{eqnarray}
{\bf J}_{5, \omega}^{a}\left({\bf k}_{1}, {\bf k}_{2}, {\bf q}
\right) 
&=&
- f_{\pi} g_{\pi NN} {g_{\omega}^{2} \over{2 M^{2}}}  
\left({{\bf q} \over{{\bf q}^{2} + m_{\pi}^{2}}} \right) 
\nonumber \\
&\times& 
\Bigg[\tau_{(1)}^{a} {\bbox{\sigma}_{(1)} \cdot {\bf k}_{2}
\over{{\bf k}_{2}^{2} + m_{\omega}^2}} +
\left(1 \leftrightarrow 2 \right)
\Bigg]~. \
\label{e:omegamec}
\end{eqnarray}
This completes the construction of the PCAC-constrained 
axial current
in models with OBEP based on the $\pi, \sigma$, and $\omega$ 
mesons. The $\rho$ meson was deliberately omitted since
its contribution is (highly) model dependent.

\section{Pion production operators}

The one-body pion production operator is well known, as well as
the two-body one associated with pion exchange. We shall 
therefore concentrate on the MECs that are associated with 
other meson ($\sigma$ and $\omega$) exchanges, as specified by 
the PCAC constraint Eq. (\ref{e:div2bm}).

\subsection{Linear sigma model}

It follows from Eq. (\ref{e:div2bm}) and the linear sigma model 
axial current Eq. (\ref{e:lsmec}) that  
\begin{eqnarray}
\Gamma_{\pi {\rm 2-b}}^{a}\left({\bf k}_{1}, {\bf k}_{2}, 
{\bf q}\right) 
&=& 
\Gamma_{\pi {\rm I}}^{a}\left({\bf k}_{1}, {\bf k}_{2}, 
{\bf q}\right) 
+
\Gamma_{\pi {\rm II}}^{a}\left({\bf k}_{1}, {\bf k}_{2}, 
{\bf q}\right) 
+
\Gamma_{\pi {\rm III}}^{a}\left({\bf k}_{1}, {\bf k}_{2}, 
{\bf q}\right) 
\nonumber \\
&=&
i {g_{0}^{3} \over{2M^{2}}} \tau_{(2)}^{a} 
\Bigg[
{\bbox{\sigma}_{(2)} \cdot {\bf k}_{2} 
\over{{\bf k}_{2}^{2} + m_{\pi}^2}} +
{\bbox{\sigma}_{(2)} \cdot {\bf q} 
\over{{\bf k}_{1}^{2} + m_{\sigma}^2}} \Bigg]
\nonumber \\
&-& 
i g_{\sigma \pi \pi} {g_{0}^{2} \over 2 M} \tau_{(2)}^{a} 
{
\bbox{\sigma}_{(2)} \cdot {\bf k}_{2}
\over{\left({\bf k}_{1}^{2} + m_{\sigma}^2 \right) 
\left({\bf k}_{2}^{2} + m_{\pi}^2\right)}}
+ \left(1 \leftrightarrow 2 \right)~, \
\label{e:pimecl}
\end{eqnarray}
as the complete linear sigma model pion production
operator to this order in $1/M$. This result corresponds to
Figs. \ref{f:1}(b), \ref{f:2}, \ref{f:3}. Although it has long
been known that the corresponding covariant amplitude is
chirally symmetric, we are not aware of anyone having used the
MEC (\ref{e:pimecl}) in nuclear physics. 

\subsection{Heterotic sigma model}

As stated before, the main difference between the linear and the 
heterotic sigma models is the axial coupling constant $g_A$,
which is induced by the new derivative-coupled interactions.
The latter, in turn, renormalize one old graph, Fig. \ref{f:2}, 
and creates two new elementary diagrams: Fig. \ref{f:3}, and 
Fig. \ref{f:1}(b)
\footnote{
Scalar meson exchange currents have been introduced into the 
analysis of $pp \rightarrow \pi^{0}pp$
on an {\it ad hoc} basis \cite{hor93,riska93}. Those papers 
are different from ours in that they do not include the
pion-sigma-exchange graph, nor was there any concern shown for 
the consistency between the nuclear wave functions and the pion 
production operators.}. 
Thus one finds
\begin{eqnarray}
\Gamma_{\pi {\rm 2-b}}^{a}\left({\bf k}_{1}, {\bf k}_{2}, 
{\bf q}; h\right) 
&=& 
\Gamma_{\pi {\rm I}}^{a}\left({\bf k}_{1}, {\bf k}_{2}, 
{\bf q}; h \right) 
+
\Gamma_{\pi {\rm II}}^{a}\left({\bf k}_{1}, {\bf k}_{2}, 
{\bf q}; h\right) 
\nonumber \\
&+&
\Gamma_{\pi {\rm III}}^{a}\left({\bf k}_{1}, {\bf k}_{2}, 
{\bf q}; h\right) 
+
\Gamma_{\pi {\rm IV}}^{a}\left({\bf k}_{1}, {\bf k}_{2}, 
{\bf q}; h\right) 
\nonumber \\
&=&
i {g_{0}^{3} \over{2M^{2}}} \tau_{(2)}^{a} 
\Bigg[g_{A} {\bbox{\sigma}_{(2)} \cdot {\bf k}_{2} 
\over{{\bf k}_{2}^{2} + m_{\pi}^2}} +
{\bbox{\sigma}_{(2)} \cdot {\bf q} 
\over{{\bf k}_{1}^{2} + m_{\sigma}^2}} 
\nonumber \\
&+& 
\left(g_{A} - 1\right) {\bbox{\sigma}_{(2)} \cdot 
\left({\bf k}_{1} - {\bf q}\right)
\over{{\bf k}_{1}^{2} + m_{\sigma}^2}} \Bigg]
\nonumber \\
&-& 
i g_{\sigma \pi \pi} g_{A}
{g_{0}^{2} \over 2 M} \tau_{(2)}^{a} 
{\bbox{\sigma}_{(2)} \cdot {\bf k}_{2}
\over{\left({\bf k}_{1}^{2} + m_{\sigma}^2\right) 
\left({\bf k}_{2}^{2} + m_{\pi}^2\right)}}
\nonumber \\ 
&+& 
 \left(1 \leftrightarrow 2 \right)~. \
\label{e:pimech}
\end{eqnarray}
This is the complete two-body pion-production operator in the
heterotic sigma model, though with all ${\cal O}(1/M^3)$ 
pion-exchange two-body operators excluded. This operator is 
only to be used with
nuclear wave functions that are solutions to the Schr\" odinger
equation with the aforementioned one pion + sigma exchange NN 
potential, Eq. (\ref{e:hpot}). 

\subsection{Omega meson exchange}

The omega-exchange $\pi$-production operator, Fig. \ref{f:4}, 
to this order in $1/M$ is
\begin{eqnarray}
\Gamma_{\pi \omega}^{a}\left({\bf k}_{1}, {\bf k}_{2}, 
{\bf q}\right) 
&=& 
- i g_{\omega}^{2}{g_{\pi NN} \over{2M}} 
\left(\tau_{(1)}^{a}  
{\bbox{\sigma}_{(1)} \cdot {\bf k}_{2} 
\over{{\bf k}_{2}^{2} + m_{\omega}^2}}
+ \left(1 \leftrightarrow 2 \right)\right)~. \
\label{e:omegapimec}
\end{eqnarray}
Note the utter functional dissimilarity between this and the two
$\sigma$-MECs (\ref{e:pimecl}), and (\ref{e:pimech}), despite 
the fact that both are due to terms in
the NN potential that are indistinguishable to this order in
nonrelativistic expansion, {\it viz.} the
isoscalar one-vector- and the scalar-meson 
exchange potentials in Eqs. (\ref{e:omegapot}), and 
(\ref{e:hpot}) respectively. Specifying the Lorentz
structure of the NN potential distinguishes the vector- from the
scalar-exchange induced terms, but even then the ambiguity 
between the two kinds of scalar MECs is survives. This
fact brings into focus the intrinsic and apparently intransigent 
ambiguities
associated with attempts to construct consistent axial MECs 
starting from the NN potential \cite{riskab92}. At this stage
the only formalism that allows systematic construction of
consistent is based on relativistic chiral Lagrangians.

We believe the preceding results to be important for 
two reasons: (i) They provide several explicit examples 
of how PCAC taken as an underlying principle leads to 
classification and construction
of admissible (PCAC-consistent) approximations to nuclear axial
current matrix elements in chiral models.
(ii) They specify PCAC-consistent, apparently novel pion 
production operators in several OBEP nuclear models.
Besides the $pp \rightarrow \pi^{0}pp$ reaction, \cite{sc97}, 
these results ought to also be directly applicable to studies 
of the $pp \rightarrow \pi^{+}d$ reaction \cite{canton98}
\footnote{
This includes the nuclear chiral perturbation theory 
($N\chi PT$), which is in the present OBE approximation just
the one-pion-exchange term that was previously treated in Refs.
\cite{oht79,sauer92}}.

\section{Summary and conclusions}

In summary, in this paper we have used PCAC to relate the 
nuclear axial current and pion production operators to each
other, and to the two-nucleon potential, in nuclear theories 
based on the nonrelativistic Schr\" odinger equation. We focused
in particular on the axial- and pion-production MECs related to
the exchange of the lightest isoscalar mesons with 
$J^{P} = 0^{+}, 1^{-}$, 
i.e., to $\sigma, \omega$ mesons. [Pions have been treated
elsewhere \cite{oht79,sauer92}, and the $\rho$ meson 
contributions are highly model-dependent.] 
We constructed axial currents and pion production operators in 
two variations of the linear sigma model, with $g_{A} = 1$, or 
$g_{A} = 1.26$, with or without $\omega$-exchange and
showed explicitly that they satisfy the PCAC constraints. All 
of these results are new, to our knowledge.

In the process we also made several assumptions and approximations:

(i) we neglected the pion-pole term in the axial charge. This is
justifiable since no need arose for it in our analysis.

(ii) we neglected all retardation effects, as well as the recoil
MECs.

(iii) we neglected all isovector-vector and/or axial vector 
meson exchanges.

(iv) we did not include nucleon or meson form factors, either 
electro-weak or strong.

Another objection, perhaps of a more theoretical 
nature, can be raised against the present calculation: 
to talk about meson production in a NR potential theory 
(where such meson d.o.f. have been ``integrated out") 
seems self-contradictory. 
We have side-stepped this problem by 
defining the pion production operator as the 
${\cal O}(f_{\pi} m_{\pi}^{2})$ term in the
divergence of the axial current Eq. (\ref{e:divme}), in
analogy with the relativistic PCAC result.
This issue can be addressed more deeply within the 
Fukuda-Sawada-Taketani-Okubo-Nishijima (FSTON) method for
constructing effective nuclear theories
\cite{sato97}, and we are currently working on it.

Partial conservation of the axial current ought to be an 
important criterion in the construction of both nuclear
two-body, or meson-exchange axial current operators, and of
nuclear two-body pion-production operators.
The latter arrise from two sources:
(i) ``elementary" meson-nucleon vertices present in the chiral
Lagrangian; and (ii) ``effective" meson-nucleon vertices due 
to the time-ordered Z-graphs, also known as pair currents.
They have not, to our knowledge, been examined from the 
present viewpoint, with the exception of Refs. 
\cite{oht79,sauer92} on the one-pion-exchange axial currents and 
the early work by Blin-Stoyle and Tint \cite{bst67} on nuclear 
pion production. Our study is only the first step in joining
these two ideas and extending them to include  
light isoscalar mesons $(\sigma, \omega)$. 

This work is based on
two fundamental assumptions: (i) quantum mechanics; and (i) PCAC.
Although our examples were drawn from two specific 
sigma models, they exemplify a far more general relation between 
the nuclear Hamiltonian and the nuclear pion production operator.
Indeed 
such a relation must hold in any calculation based on the two
aforementioned principles, and in nuclear chiral perturbation
theory in particular. This relationship is almost completely 
unexplored in the last mentioned setting, a situation that must 
be remedied.

{\it Note added in proof.} We have learnt of the related work by
S. M. Ananyan, Phys. Rev. C {\bf 57}, 2669 (1998) only
after the present paper had been submitted.

\acknowledgments

The authors would like to thank Kuniharu Kubodera and Fred Myhrer 
for valuable discussions and comments on the manuscript, as well 
as for providing an opportunity for this collaboration to take 
place. We also wish to thank Mr. T. Nogi for help in checking
the calculations.

\appendix
\section{Axial charge density in the heterotic sigma model}
\label{app}

The heterotic sigma model is a 
chirally symmetric field theoretic model that leads to 
an axial current with arbitrary $g_A (\neq 1)$, and a mixture of 
pseudoscalar and pseudovector pion-nucleon couplings \cite{ax96}.
The Lagrangian density of this model is given by
\begin{eqnarray}
{\cal L} &=& \bar{\psi}{\rm i}{\partial{\mkern-10mu}{/}}\psi -
g_{0}\bar{\psi}\left[\sigma + 
{\rm i}\gamma_{5} \bbox{\pi} \cdot \tau\right]\psi  
+ {1 \over 2} \left(\partial_{\mu} 
\bbox{\phi}\right)^{2} - V(\bbox{\phi}^{2}) 
\nonumber \\
&+& \left({g_{A} - 1 \over{f_{\pi}^{2}}}\right) \left[
\left(\bar{\psi}\gamma_{\mu} {\bbox{\tau}\over 2}\psi \right) \cdot
\left(\bbox{\pi} \times \partial^{\mu} \bbox{\pi}\right)
+ \left(\bar{\psi}\gamma_{\mu} \gamma_{5} {\bbox{\tau}\over 2}\psi 
\right) \cdot
\left(\sigma \partial^{\mu} 
\bbox{\pi} - \bbox{\pi} \partial^{\mu} \sigma \right) \right] ~,
\label{e:lag3} \
\end{eqnarray}
where $\bbox{\phi} = (\sigma,\bbox{\pi})$
is a column vector and $V$ is the same potential 
as in the linear sigma model.
The (partially conserved) axial-vector Noether current in this 
model reads
\begin{eqnarray}
{\bf J}_{\mu 5}^{a} &=& 
\left(\bar{\psi}\gamma_{\mu} \gamma_{5} {\bbox{\tau}\over 2}\psi 
\right)^{a} -
\left(\bbox{\pi} \partial_{\mu} \sigma - \sigma \partial_{\mu} 
\bbox{\pi}\right)^{a}
\nonumber \\
&+& 
\left({g_{A} - 1 \over{f_{\pi}^{2}}}\right) \Bigg[
\left(\bar{\psi}\gamma_{\mu} \gamma_{5} 
{\bbox{\tau}\over 2}\psi 
\cdot \bbox{\pi} \right) \bbox{\pi}^{a} 
\nonumber \\
&+& 
\sigma^{2}
\left(\bar{\psi}\gamma_{\mu} \gamma_{5} 
{\bbox{\tau}\over 2}\psi \right)^{a} 
+ \sigma
\left(\bar{\psi} \gamma_{\mu} {\bbox{\tau}\over 2}\psi 
\times \bbox{\pi} \right)^{a} 
\Bigg]  ~,
\label{e:axi4} \
\end{eqnarray}
After shifting the sigma field this can be written as
\begin{eqnarray}
{\bf J}_{\mu 5}^{a} &=& g_{A}
\left(\bar{\psi}\gamma_{\mu} \gamma_{5} {\bbox{\tau}\over 2}\psi 
\right) 
+ f_{\pi} \partial_{\mu} \bbox{\pi}
+ \left(s \partial_{\mu} \bbox{\pi} 
- \bbox{\pi} \partial_{\mu} s \right)^{a} 
\nonumber \\
&+& 
\left({g_{A} - 1 \over{f_{\pi}^{2}}}\right) \Bigg[
\left(\bar{\psi}\gamma_{\mu} \gamma_{5} 
{\bbox{\tau}\over 2}\psi 
\cdot \bbox{\pi} \right) \bbox{\pi}^{a} 
\nonumber \\
&+&  
s \left(2 f_{\pi} + s \right)
\left(\bar{\psi}\gamma_{\mu} \gamma_{5} 
{\bbox{\tau}\over 2}\psi \right)^{a} 
+ \left(f_{\pi} + s \right)
\left(\bar{\psi} \gamma_{\mu} {\bbox{\tau}\over 2}\psi 
\times \bbox{\pi} \right)^{a} 
\Bigg]  ~.
\label{e:axi5} \
\end{eqnarray}
Thus we see that the nucleon axial current has acquired a new
coupling constant: $g_{A} \neq 1$, which was the purpose of this 
model.

The (new) derivative coupling terms in the Lagrangian
(\ref{e:lag3}) modify the canonical momenta as follows
\begin{eqnarray}
\Pi_{\sigma}  &=& \dot{\sigma} - 
\left({g_{A} - 1 \over{f_{\pi}^{2}}}\right) 
\left(\psi^{\dagger} \gamma_{5}
{\bbox{\tau} \cdot \bbox{\pi}\over 2}\psi \right) 
\\
\Pi_{\pi}^{a} &=& \dot{\bbox{\pi}^{a}} + 
\left({g_{A} - 1 \over{f_{\pi}^{2}}}\right) 
\left[\left(\psi^{\dagger}
{\bbox{\tau} \times \bbox{\pi}\over 2}\psi \right)^{a} + 
\sigma \psi^{\dagger} \gamma_{5} 
{\bbox{\tau}^{a}\over 2}\psi \right]  ~.
\label{e:cm} \
\end{eqnarray}
We see that the axial charge retains its linear sigma model
form when written out in terms of canonical fields and their
associated momenta:
\begin{eqnarray}
\rho_{5}^{a} &=& {\bf J}_{0 5}^{a} =
\psi^{\dagger} \gamma_{5} {\bbox{\tau}^{a}\over 2}\psi 
- \left(\bbox{\pi}^{a} \Pi_{\sigma} - 
\sigma \bbox{\Pi}_{\pi}^{a} \right) ~.
\label{e:axi6} \
\end{eqnarray}
Hence we see that the axial charge carried by the nucleon is 
unchanged as compared with
the one in the linear sigma model, i.e., we have $g_A = 1$ here,
which was to be proven.

\end{document}